\documentclass[]{emulateapj}
\usepackage{amsmath}
\begin{document}

\title{Modelling the radio emission from Cyg OB2 \#5: a quadruple system?}

\author{M. Kennedy} 
\affil{University of Victoria, Department of Physics and Astronomy, 
       3800 Finnerty Rd, Victoria, BC, Canada V8P 5C2} 
\email{mgk@uvic.ca}
\author{S.M. Dougherty\altaffilmark{1} and A. Fink} 
\affil{National Research Council Herzberg Institute for Astrophysics 
       Dominion Radio Astrophysical Observatory,\\ 
       P.O. Box 248, Penticton, BC, Canada V2A 6J9}
\email{sean.dougherty@nrc.ca, amy.fink@nrc.ca}
\altaffiltext{1}{University of Calgary, 
      Department of Physics and Astronomy,\\ 
      2500 University Dr. N.W., Calgary, AB, Canada T2N~1N4}
\and 
\author{P.M. Williams} 
\affil{Institute for Astronomy, Royal Observatory, 
       Blackford Hill, Edinburgh, Scotland EH9 3HJ}
\email{pmw@roe.ac.uk}

\begin{abstract}
Fifty observations at frequencies between 1.4 GHz and 43 GHz of the
6.6-day O6.5-7+O5.5-6 binary \object{Cyg OB2 \#5} using the Very Large
Array over 20 years are re-examined. The aim is to determine the
location and character of the previously detected variable radio
emission. The radio emission from the system consists of a primary
component that is associated with the binary, and a non-thermal source
(NE), $0.8\arcsec$ to the NE of the binary that has been ascribed to a
wind-collision region (WCR) between the stellar winds of the binary
and that of a B-type star (Star D) to the NE .  Previous studies have
not accounted for the potential contribution of NE to the total radio
emission, most especially in observations where the primary and NE
sources are not resolved as separate sources. NE shows no evidence of
variation in 23 epochs where it is resolved separately from the
primary radio component, demonstrating that the variable emission
arises in the primary component. Since NE is non-variable, the radio
flux from the primary can now be well determined for the first time,
most especially in observations that do not resolve both the primary
and NE components. The variable radio emission from the primary
component has a period of $6.7\pm0.3$~years which is described by a
simple model of a non-thermal source orbiting within the stellar wind
envelope of the binary. Such a model implies the presence of a third,
unresolved stellar companion (Star C) orbiting the 6.6-day binary with
a period of 6.7 years and independent of Star D to the NE. The
variable non-thermal emission arises from either a WCR between Star C
and the binary system, or possibly from Star C directly.  The model
gives a mass-loss rate of $3.4\times10^{-5}$~M$_\odot$\,yr$^{-1}$ for
Cyg OB2 \#5, unusually high for an Of supergiant and comparable to
that of WR stars, and consistent with an unusually strong He\,{\sc i}
1.083-$\mu$m emission line, also redolent of WR stars. An examination
of radial velocity observations available from the literature suggests
reflex motion of the binary due to Star C, for which a mass of
$23^{+22}_{-14}$~M$_{\odot}$ is deduced.  The natures of NE and Star D
are also examined. If NE is a WCR, as suggested by other authors, then
the required mass-loss rate is an order of magnitude higher than
expected for an early B-type dwarf, and only just consistent with a
supergiant. This raises the question of NE as a WCR, but its
non-thermal luminosity is consistent with a WCR and a comparison of
reddening between \object{Cyg OB2 \#5} and Star D do not rule out an
association, implying \object{Cyg OB2 \#5} is a quadruple
system. Pursuing alternative models for NE, such as an unassociated
background source, would require very challenging observations.
\end{abstract}

\section{Introduction}
\object{Cyg OB2 \#5} (\object{V729 Cyg}, \object{BD $+40\,4220$}) is
an eclipsing binary system consisting of two O-type supergiants
orbiting in a 6.6-day period \citep{Hall:1974, Leung:1978,
Rauw:1999}. This system is one of several luminous O-star systems in
the Cyg OB2 association that shows evidence of variable radio emission
\citep{Persi:1983, Persi:1990, Bieging:1989} in observations gathered
over $\sim20$~yrs. The radio emission appears to have two states: a
low-flux state of $\sim2$~mJy at 4.8 GHz where the spectral index is
consistent with thermal emission from a stellar wind, and a high-flux
state of $\sim8$~mJy at 4.8 GHz, where the spectral index is flatter
than in the low state.  The variations appear to have a $\sim7$-year
period \citep{Miralles:1994} and have been attributed to variable
non-thermal emission from an expanding plasmon arising in the binary
\citep{Bieging:1989, Persi:1990, Miralles:1994}.

Observations by \citet{Abbott:1981} with the VLA revealed two radio
components: a primary component associated with the binary and a
secondary radio source (hereafter NE) $\sim0.8\arcsec$ to the
NE of the primary radio source. \citet{Miralles:1994} confirmed the
existence of NE and a $3-6$~cm spectral index of $-0.5\pm0.3$ hinted
at non-thermal emission. \citet{Contreras:1997} argue that this
emission is the result of a wind-collision region (WCR) between the
stellar wind of the binary and that of a B0-2 V star (hereafter Star
D), $0.9\arcsec$ from the binary to the NE \citep{Contreras:1997}.

All previous analyses of the radio emission from \object{Cyg OB2 \#5}
are based on observations from the Very Large Array (VLA), obtained in
all the different configurations of the array, and hence data covering
different spatial frequency (i.e. baseline length) ranges.  Dependent
on the spatial frequency coverage, the observations may or may not
resolve the emission from both the primary and NE components. None of
the previous analyses in the literature have attempted to take this
into account.  Furthermore, no attempt has been made to determine
either the continuum spectrum over a broad wavelength range or the
location of the variable emission. As a result, not only is the nature
of the radio emission from NE unknown, the evolution of the radio
emission of the primary remains uncertain, which may account for the
poor fits of models of the non-thermal emission to the observables
\cite[e.g.][]{Miralles:1994}.

In this paper all VLA archive radio observations of \object{Cyg OB2
\#5} are re-examined to produce a consistently calibrated data
set. The analysis accounts for the changes in the resolution of the
array and presents a consistent treatment of the emission from both
the primary and NE sources in each observation.  For the first time,
the nature of the emission of both the primary and NE can be
determined throughout the $\sim20$ years of the observations.

\section{Observations}
\subsection{Very Large Array}

A total of 50 VLA observations of \object{Cyg OB2 \#5} obtained
between 1983 and 2003 were extracted from the NRAO archive.  These
data comprise 30 observations at 4.8~GHz (C band), 17 observations at
8.4~GHz (X band), 2 observations at 14.9~GHz (U band), and a single
observation at 43.3~GHz (Q band). Examination of the few VLA
observations at 1.4 GHz (L band) obtained in the low-resolution, C and
D configurations were contaminated by extended low-surface brightness
emission and were not incorporated in this study.

\begin{deluxetable}{cccccc}
\tablewidth{0pt}
\tabletypesize{\scriptsize}
\tablecaption{VLA Observations.\label{table:VLA}}
\tablehead{\colhead{Date}&\colhead{MJD}	&\colhead{Cfg.}& \colhead{Phase}&\colhead{Primary}&\colhead{NE}\\
\colhead{}&\colhead{}&\colhead{} &\colhead{Calibrator}&\colhead{Component}&\colhead{Component}\\
\colhead{}&\colhead{}&\colhead{} &\colhead{Flux}&\colhead{Flux}&\colhead{Flux}\\
\colhead{(Y/M/D)}&\colhead{}&\colhead{}&\colhead{(Jy)} &\colhead{(mJy)} &\colhead{(mJy)}}
\startdata
\tableline
\tableline
\multicolumn{6}{c}{1.4GHz}\\
\tableline
\tableline
87/10/16& 47084 &AB &\phm{9,9}5.17$^{a,1}$ &\multicolumn{2}{c}{$2.98\pm0.13$\tablenotemark{b}} \\
87/11/09& 47108 &AB &\phn5.17$^1$ &\multicolumn{2}{c}{$3.02\pm0.14$} \\
\tableline
\tableline
\multicolumn{6}{c}{4.8GHz}\\
\tableline
\tableline
83/08/18&45564	&A		&4.73	&$6.21\pm0.19$	&$0.94\pm0.10$\\
84/05/23&45843 	&C		&4.23	& \multicolumn{2}{c}{$8.15\pm0.24$}\\
84/05/27&45847	&C		&\phn3.02$^2$&\multicolumn{2}{c}{$8.05\pm0.26$}  \\
84/09/06&45949	&D		&4.41	&\multicolumn{2}{c}{$6.81\pm0.20$}  \\
84/09/15&45958	&D		&4.45	&\multicolumn{2}{c}{$7.67\pm0.23$}  \\
84/09/20&45963	&D		&4.34	&\multicolumn{2}{c}{$7.28\pm0.22$}  \\
84/09/22&45965 	&D		&4.31	&\multicolumn{2}{c}{$7.55\pm0.23$}  \\
84/09/24&45967	&D		&4.27	&\multicolumn{2}{c}{$7.13\pm0.21$}  \\
84/09/28&45971	&D		&4.20	&\multicolumn{2}{c}{$6.63\pm0.20$}  \\
84/10/16&45989	&D		&\phn2.94$^2$&\multicolumn{2}{c}{$8.41\pm0.25$} \\
85/11/21&46309	&D		&4.04	&\multicolumn{2}{c}{$5.37\pm0.16$}  \\
86/01/13&46443	&D		&4.14	&\multicolumn{2}{c}{$4.18\pm0.13$}  \\
87/06/04&46952	&D		&3.56	&\multicolumn{2}{c}{$3.75\pm0.16$}  \\
87/06/06&47196	&D		&3.60	&\multicolumn{2}{c}{$3.90\pm0.27$}  \\
87/10/16&47084	&AB		&\phn3.10$^1$&$2.78\pm0.10$	&$0.71\pm0.10$	\\
87/11/09&47108	&AB		&\phn3.10$^1$&$2.57\pm0.10$	&$0.72\pm0.10$	\\
88/02/05&47196	&B		&\phn3.10$^1$&\multicolumn{2}{c}{$4.15\pm0.22$}\\
89/09/18&47787	&BC		&2.76	&\multicolumn{2}{c}{$7.67\pm0.23$}  \\
92/11/02&48928	&A		&2.92	&$2.94\pm0.10$	&$0.93\pm0.10$	\\
92/12/18&48974	&A		&3.01	&$2.70\pm0.10$	&$0.74\pm0.10$	\\
92/12/19&48975	&A		&3.02	&$2.76\pm0.10$	&$0.73\pm0.10$	\\
93/05/01&49108	&B		&3.06	&$2.56\pm0.10$	&$0.85\pm0.10$	\\
93/12/27&49348	&D		&3.02	&\multicolumn{2}{c}{$3.67\pm0.24$}\\
94/04/09&49451	&A		&2.90	&$2.53\pm0.10$	&$0.94\pm0.10$	\\
94/04/16&49458	&A		&2.90	&$2.37\pm0.10$	&$0.84\pm0.10$	\\
95/04/27&49834	&D		&3.00	&\multicolumn{2}{c}{$5.81\pm0.24$}\\
95/06/10&49878	&AD		&3.15	&$5.13\pm0.15$	&$0.89\pm0.10$	\\
95/06/12&49880	&AD		&3.14	&$5.16\pm0.15$	&$0.64\pm0.10$	\\
97/01/04&50452	&A		&3.08	&$7.21\pm0.22$	&$0.96\pm0.10$	\\
00/06/30&51725	&CD		&\phn1.94$^3$&\multicolumn{2}{c}{$3.30\pm0.10$}\\
\tableline\tableline
\multicolumn{6}{c}{8.4GHz}\\
\tableline
\tableline
90/06/01&48043	&AB	&3.05	&$8.07\pm0.24$	&$0.57\pm0.10$	\\
91/10/03&48532	&AB	&3.18	&$9.03\pm0.27$	&$0.40\pm0.10$	\\
92/07/16&48819	&D	&3.18	&\multicolumn{2}{c}{$6.99\pm0.21$}\\
92/11/02&48928	&A	&3.15	&$4.27\pm0.13$	&$0.66\pm0.10$	\\
92/12/19&48975	&A	&3.25	&$3.99\pm0.12$	&$0.57\pm0.10$	\\
93/05/01&49108	&B	&3.23	&$4.11\pm0.12$	&$0.39\pm0.10$	\\
93/10/31&49291	&D	&3.31	&\multicolumn{2}{c}{$4.30\pm0.25$}\\
93/12/27&49348	&D	&3.10	&\multicolumn{2}{c}{$4.90\pm0.30$}\\
94/04/09&49451	&A	&2.86	&$3.48\pm0.10$	&$0.52\pm0.10$	\\
94/04/16&49458	&A	&2.90	&$3.39\pm0.10$	&$0.65\pm0.10$	\\
95/04/27&49834	&D	&2.80	&\multicolumn{2}{c}{$7.60\pm0.23$}\\
96/02/02&50115	&BC	&2.88	&\multicolumn{2}{c}{$7.84\pm0.24$}\\
96/12/28&50445	&A	&2.91	&$8.08\pm0.24$	&$0.62\pm0.10$	\\
98/07/17&51011	&B	&2.61	&$9.02\pm0.27$	&$0.42\pm0.10$	\\
00/06/30&51725	&CD	&\phn2.66$^3$&\multicolumn{2}{c}{$3.82\pm0.16$}\\
03/06/06&52796	&A	&3.01	&$7.17\pm0.22$	&$0.36\pm0.10$	\\
03/09/09&52891	&A	&3.19	&$6.09\pm0.18$	&$0.37\pm0.10$	\\
\tableline
\tableline
\multicolumn{6}{c}{14.9GHz}\\
\tableline
\tableline
95/04/27&49834	&D	&9.49	&\multicolumn{2}{c}{$9.39\pm0.47$}\\
00/06/30&51725	&CD	&\phn2.87$^3$&\multicolumn{2}{c}{$5.27\pm0.26$}\\
\tableline
\tableline
\multicolumn{6}{c}{43.3GHz}\\
\tableline
\tableline
96/12/29&50446	&A	&1.13	&\multicolumn{2}{c}{$12.17\pm0.61$}\\
\enddata
\tablenotetext{a}{Name of the phase-reference calibrator used if not
  B2005+403 ($\equiv$J2004+404). 1=B2050+364; 2=B2200+420; 3=B2013+370. }
\tablenotetext{b}{Single values given for the flux are the total
    flux of emission from both the primary and secondary components. }
\end{deluxetable}

These data were edited and calibrated by a standard approach using the
NRAO's {\sc aips} software package.  Each observation was first
examined to remove bad data. Phase-only antenna gain solutions were
then established for all calibrators by self-calibration. Antennas
showing random fluctuations in phase were examined closely, with
further data editing as necessary. Both amplitude and phase
calibration solutions were then derived for the calibrators.  Again,
further editing was applied as necessary. The absolute flux scale was
established through bootstrapping the amplitude of the primary
calibrator (either 3C286 or 3C84) to the phase calibrator, typically
B2005+403 ($\equiv$J2004+404), giving a flux uncertainty typically
determined to be
$\sim3-5$\%\footnote{\url{http://www.vla.nrao.edu/astro/calib/manual/}}.
Finally, phase and amplitude solutions of the phase calibrator were
interpolated across the observations of \object{Cyg OB2 \#5}. A final
examination of the calibrated visibilities was made prior to image
construction.

The visibility data for \object{Cyg OB2 \#5} were deconvolved via
model fitting using the {\sc smerf} patch \citep{Reid:2006} to the
{\sc difmap} package \citep{Shepherd:1995}. This is a technique used
widely in VLBI image reconstruction, rather than the well-known {\sc
clean} technique.  An initial ``best'' model was established from
imaging the 8.4-GHz visibilities obtained with the highest resolution,
A configuration of the VLA.  This was then used as the initial model
for the remaining frequencies.

\begin{figure}[t]
\plotone{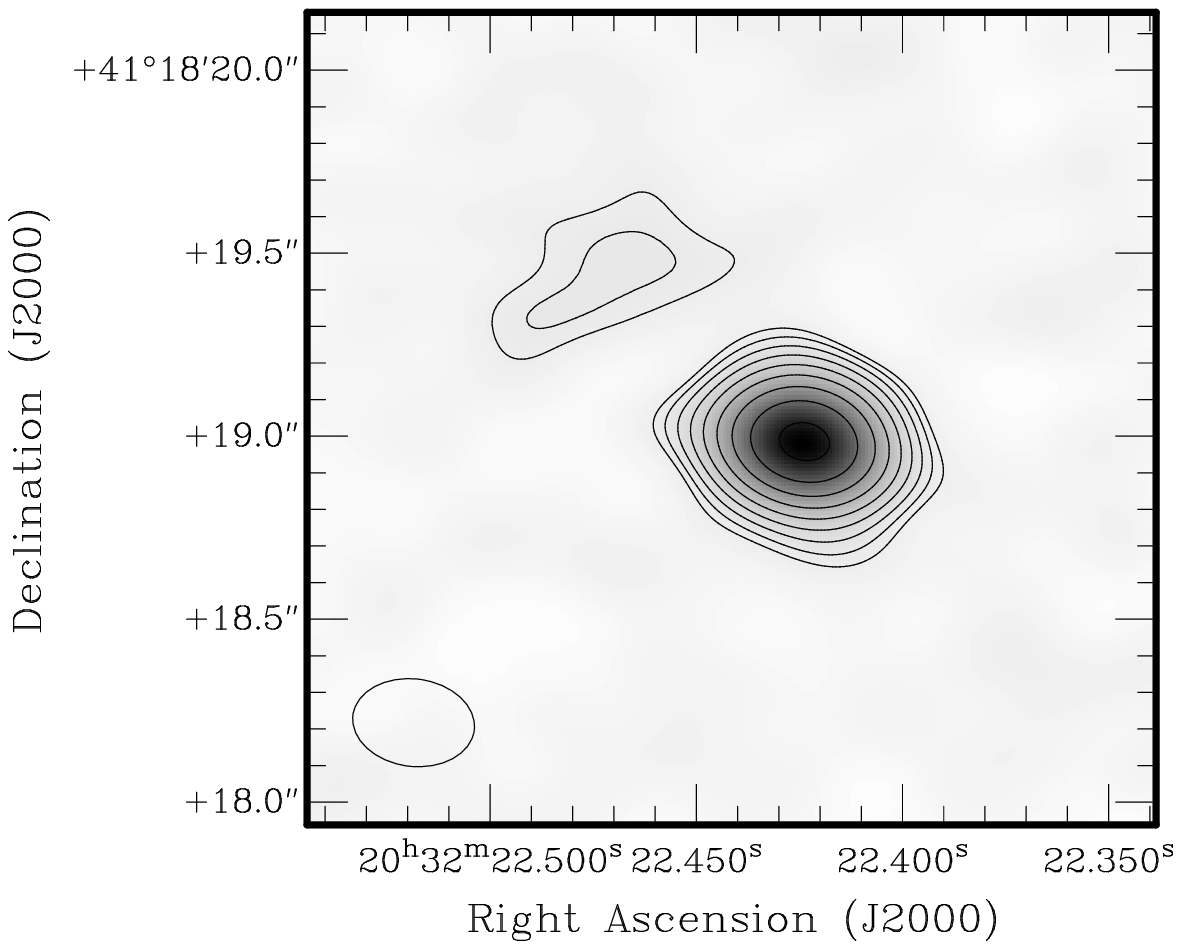}
\plotone{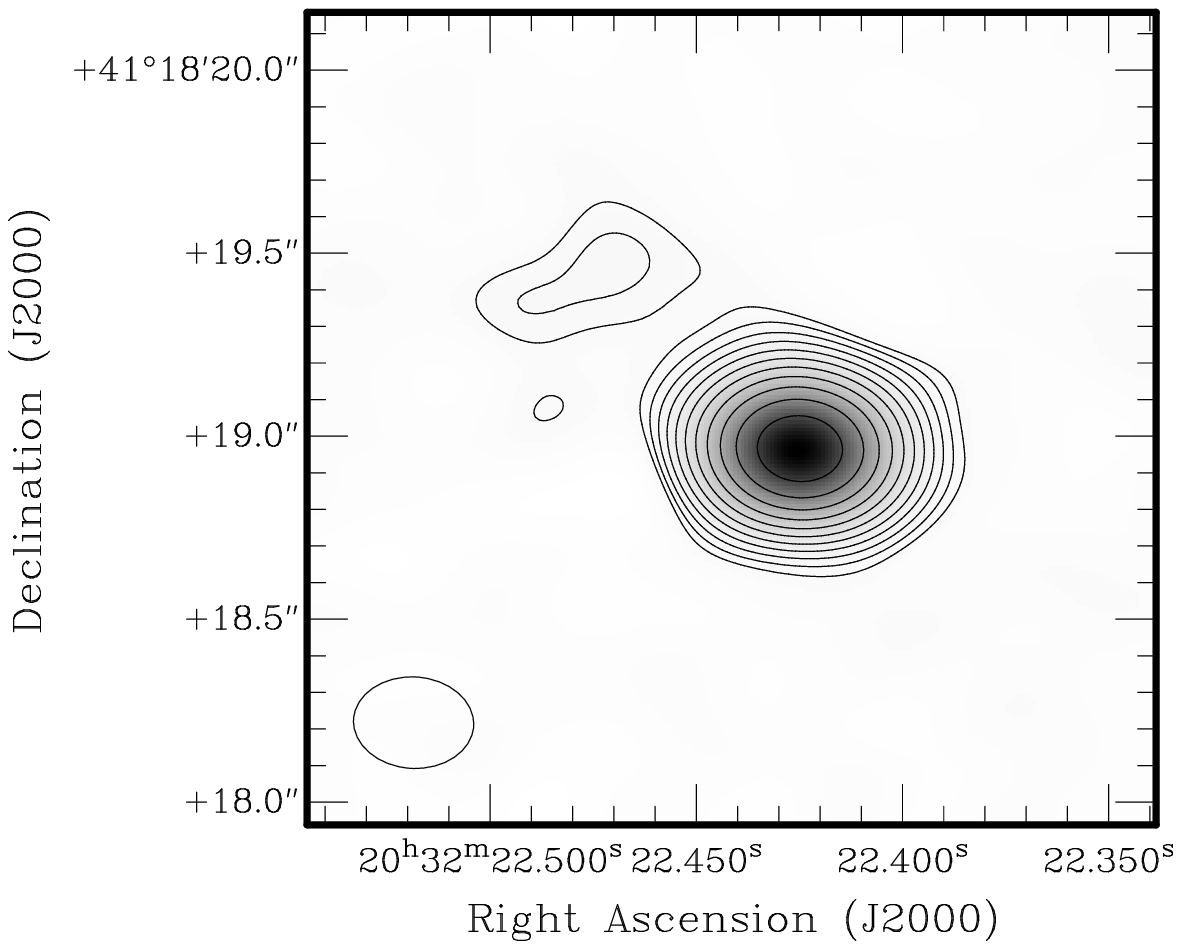}
\caption{\small Two examples of the deconvolved VLA images at 8.4 GHz
  that show the primary and NE sources. The top image is from 1992
  December 19 during a low emission state and the lower image is from
  1996 December 28 during a high emission state. From these two images
  it is seen that the flux of the primary source increases between the
  low and high flux states, whereas the flux of NE remains
  constant. The position of the primary component is consistent with
  the position of the binary system. The first contour level is
  $3\sigma$, where $\sigma$ is the image rms, which is 0.044 mJy and
  0.028 mJy in the top and bottom images respectively. The contour
  spacings represent increases by a factor of 1.5 from the $3\sigma$
  level.  The FWHM of the synthesized beam is shown in the lower left
  corner of the images.}
\label{fig:images}
\end{figure}

As noted previously, the radio emission from \object{Cyg OB2 \#5}
consists of two components: the primary source associated with the
O-star binary and NE, which is less bright than the primary. To
establish models for both these sources, the model of the primary was
first established.  Phase-only self-calibration was then applied to
further improve the antenna gain phase solutions derived initially
from phase referencing alone.  The model of NE was then
established. The final model was determined by first fixing the
location of NE and allowing the remaining model parameters to
converge, before a final model fit with all parameters free to
converge. Examples of the resulting deconvolved images at 8.4~GHz are
shown in Fig.~\ref{fig:images}.

At 8.4 GHz, the two components were readily resolved in all
observations obtained with A and B configuration of the VLA, whereas
at 4.8 GHz the two components were only resolved in A-configuration
observations. In all of these observations, NE was always detected. In
C and D configurations, the two components are not resolved separately
at any of the observing frequencies, with only a single unresolved
source being observed.

The source fluxes were taken directly from the model-fitting
parameters. The flux uncertainties at both 4.8 GHz and 8.4 GHz were
taken to be the maximum of either the rms image uncertainty or 3\% of
the source flux. At both 14.9 GHz and 43.3 GHz, the uncertainty was
taken to be the maximum of either the image rms or 5\% of the flux.  A
summary of the observations and observed fluxes is presented in Table
\ref{table:VLA}.

\begin{deluxetable}{ccccc}
\tablewidth{0pt}
\tabletypesize{\scriptsize}
\tablecaption{MERLIN Observations\label{table:MERLIN}}
\tablehead{\colhead{Date}&\colhead{MJD}&\colhead{Phase\tablenotemark{a}}&\colhead{Primary}&\colhead{Secondary}\\
\colhead{}&\colhead{}&\colhead{Calibrator}&\colhead{Component}&\colhead{Component}\\
\colhead{}&\colhead{}&\colhead{Flux}&\colhead{Flux}&\colhead{Flux} \\
\colhead{(Y/M/D)}&\colhead{}&\colhead{(Jy)}&\colhead{(mJy)}&\colhead{(mJy)}}
\startdata
\tableline
\tableline
\multicolumn{5}{c}{1.4GHz}\\
\tableline
\tableline
96/02/05&50118	&$3.18$	&$4.57\pm0.14$	&$1.43\pm0.14$\\
96/03/14&50156	&$2.56$	&$3.00\pm0.16$	&$1.33\pm0.16$\\
\tableline
\tableline
\multicolumn{5}{c}{4.8GHz}\\
\tableline
\tableline
96/11/14&50401	&$3.02$	&$5.82\pm0.13$	&$-$\tablenotemark{b}\\
\enddata
\tablenotetext{a}{Phase-reference calibrator was B2005+403
($\equiv$J2004+404) for all observations.}
\tablenotetext{b}{NE was ``resolved out''.}
\end{deluxetable}

\subsection{MERLIN}

Three epochs of MERLIN observations were extracted from the MERLIN
archive, with two observations (1996 February 5 and 1996 March 14) at
1.4 GHz (L band) and one observation (1996 November 14) at 4.8 GHz (C
band). Editing and calibration of the visibilities was performed using
{\sc aips} in a similar fashion to the VLA observations. The different
performance of the individual telescopes of the MERLIN array was
accounted for by applying different weights to each antenna according
to their
sensitivity\footnote{\url{http://www.merlin.ac.uk/user\_guide/OnlineMUG/}}.
As in the VLA observations, the observations were phase referenced
using B2005+403 and the absolute flux scale was established from
3C286.  The visibility data for \object{Cyg OB2 \#5} were deconvolved
similarly to the VLA observations through modelling with {\sc
smerf}. However, no phase-only self calibration was performed during
the modelling process due to the small number of elements in the
array.

\begin{figure}[t]
\plotone{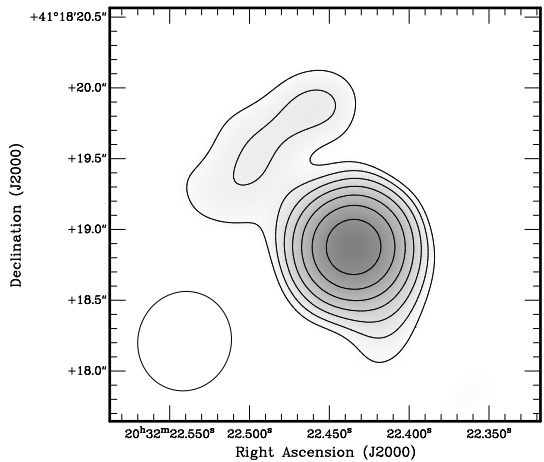}
\caption{\small MERLIN image of Cyg OB2 \#5 at 1.4 GHz from 1996 March
  14, showing that both components are resolved. A taper of
  250~k$\lambda$ was required to recover the secondary component. The
  resulting FWHM of the synthesized beam is shown in the lower left
  corner. The first contour level is $3\sigma$, where $\sigma$ is the
  image rms, which is 0.20 mJy. The contour spacings represent
  increases by a factor of 1.5 from the $3\sigma$ level. }
\label{fig:MERLIN}
\end{figure}

At 1.4 and 4.8~GHz, both NE and primary emission components are
resolved individually with MERLIN. However, NE was heavily resolved
due to the lack of low spatial frequency (i.e. short spacing) coverage
in the array. To address this issue, a Gaussian taper with HWHM of
250~k$\lambda$ was applied to the 1.4-GHz visibilities to increase the
weight of the low spatial frequency data. At 4.8 GHz, NE was
``resolved out'' and only the primary component was detected. The
primary emission model at 4.8-GHz had a major axis of 77~mas and an
axial ratio of 0.90.  A summary of the MERLIN observations is
presented in Table~\ref{table:MERLIN}. The flux uncertainties quoted
are the rms image uncertainties. The deconvolved image from 1996 March
14 is shown in Fig.~\ref{fig:MERLIN} .

\subsection{Submillimetre Observations}                          

Two epochs of 350-GHz (850~$\mu$m) photometric observations of
\object{Cyg OB2 \#5} were obtained during CANSERV time with the SCUBA
bolometric receiver system on James Clerk Maxwell Telescope
(JCMT). The first observation was taken on 1998 July 7 (MJD~51001)
with the flux calibration established from the peak flux of the
planetary map of Uranus.  The second observation was taken on 1999
June 4 (MJD~51333) with Mars as the flux calibrator. Sky dips were
used at both epochs to determine the sky opacity. Data reduction and
calibration were performed following a standard procedure using the
SCUBA User Reduction
Facility\footnote{\url{http://docs.jach.hawaii.edu/star/sun216.htx/sun216.html}}.
The observed fluxes at 850~$\mu$m were found to be $27\pm7$~mJy on
1998 July 7 and $47\pm8$~mJy on 1999 June 4. Given the sizes of
the uncertainties in these values, their difference does not provide
compelling evidence for variability. In the absence of additional
submillimetre data to suggest otherwise, it is assumed that these
radiometry data are  consistent with each other.

\subsection{Imaging and the NE Companion}
\label{sec:irobs}
\begin{figure}[t]
\plotone{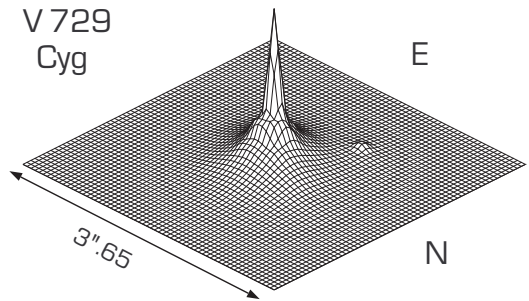}
\caption{\small Maximum-entropy reconstruction of the $2.2\mu$m image
  of V729Cyg oriented to show the companion to the NE (Star D) and
  drawn on a logarithmic intensity scale to compress the dynamic
  range.}
\label{fig:V729K}
\end{figure}

An infrared $K$-band image of \object{Cyg OB2 \#5} was observed with
the United Kingdom Infrared Telescope (UKIRT) on 1996 May 28 using the
IRCAM3 camera and 5$\times$ magnifier, giving an image scale of 0.057
arcsec~pixel$^{-1}$. It was operated in shift-and-add mode, with each
observation broken up into a large number of very short
integrations. Provided that the field contains one, dominant point
source (the central binary in this case), the centroid of this source
on the array is located for each integration and the images are
shifted in real time to bring them into alignment. Such a system is
well suited to searching for faint companions to bright point sources
and two such observations, each comprising 1000 integrations of
36-msec duration, were taken of \object{Cyg OB2 \#5}. The same setup was used
to observe HD 203856 at comparable airmass 30 minutes later to provide
a point-spread function (PSF) calibrator.  The reconstructed image is
shown in Fig.~\ref{fig:V729K}.

Fig.~\ref{fig:V729K} shows Star D to the NE, $0.90\pm0.03\arcsec$ from
the central binary at a P.A. of 55$\pm$1 $\degr$. This is consistent
with the separations from the optical images ($0.98\pm0.06\arcsec$ at
P.A. $61\pm7\degr$ from CCD images and $0.948\pm0.043\arcsec$ at
P.A. $54\pm4\degr$ from Hipparcos) given by
\citet{Contreras:1997}. The infrared magnitude difference between the
binary and companion was measured to be $\Delta K = 3.1\pm0.1$.  From
photometric observations made with the Carlos S\'{a}nchez Telescope
(Tenerife) in 1997, it is deduced that the system magnitude would
have been $K=4.52$ at the time of the imaging observation (phase
0.08 on the photometric elements of \cite{Linder:2009}), giving
$K=7.6\pm0.1$ for Star D.

If Star D is associated with the central binary, it is expected to
have the same reddening \citep[$A_K = 0.7$, deduced from $A_V
=6.4$][]{Torres-Dodgen:1991}, and distance modulus
\citep[10.7,][]{Torres-Dodgen:1991} as \object{Cyg OB2 \#5}, giving an
absolute magnitude $M_K = -3.8$. This is $\sim 0.5$ magnitude too
luminous for a B0~V star (cf. \citet{Vacca:1996} with intrinsic
$(V-K)$ from \citet{Ducati:2001}), so it is suggested here that Star D
is more evolved than a main sequence star, but probably not a
giant. The luminosity class of Star D has been inferred previously
from photometry, yet spectroscopy is necessary to determine both the
stellar type and luminosity class.

\section{Analysis}

\subsection{Variations in Radio Emission}

\begin{figure}[t]
\plotone{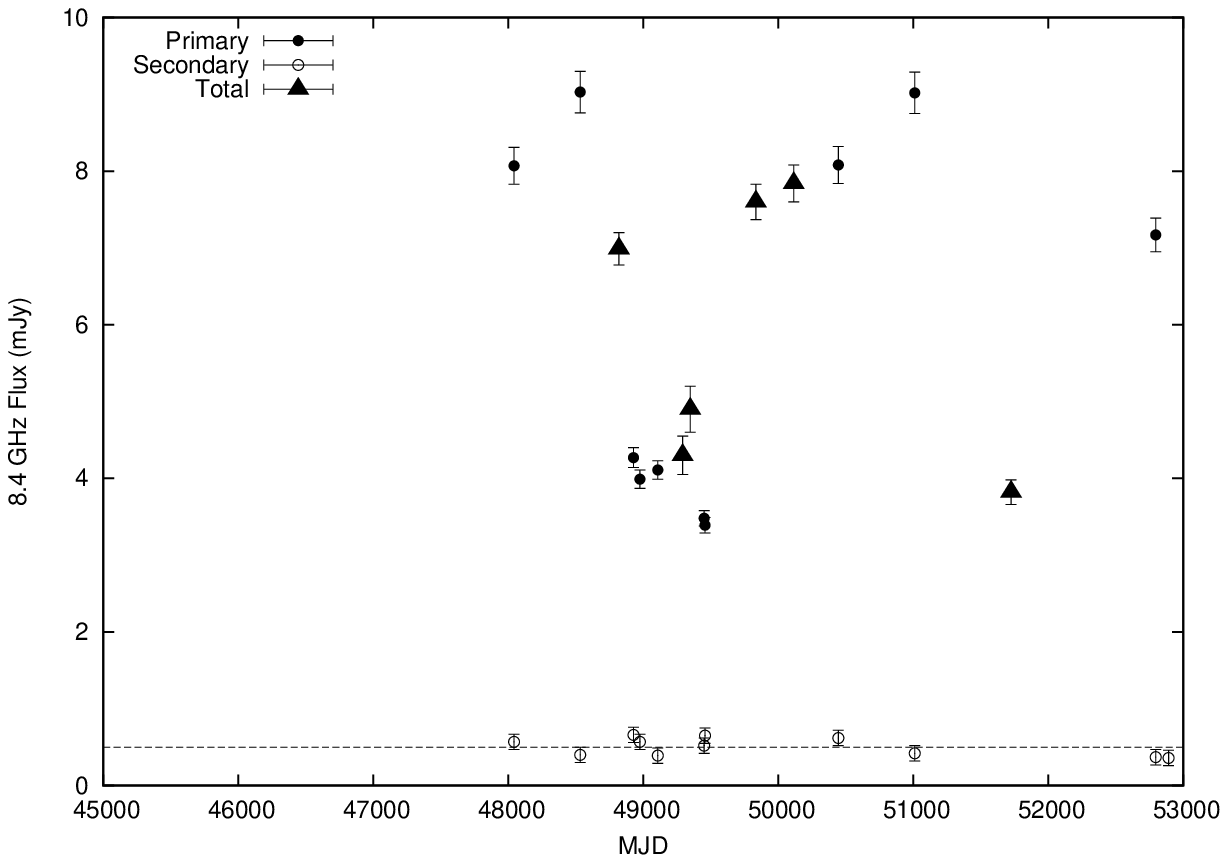}
\plotone{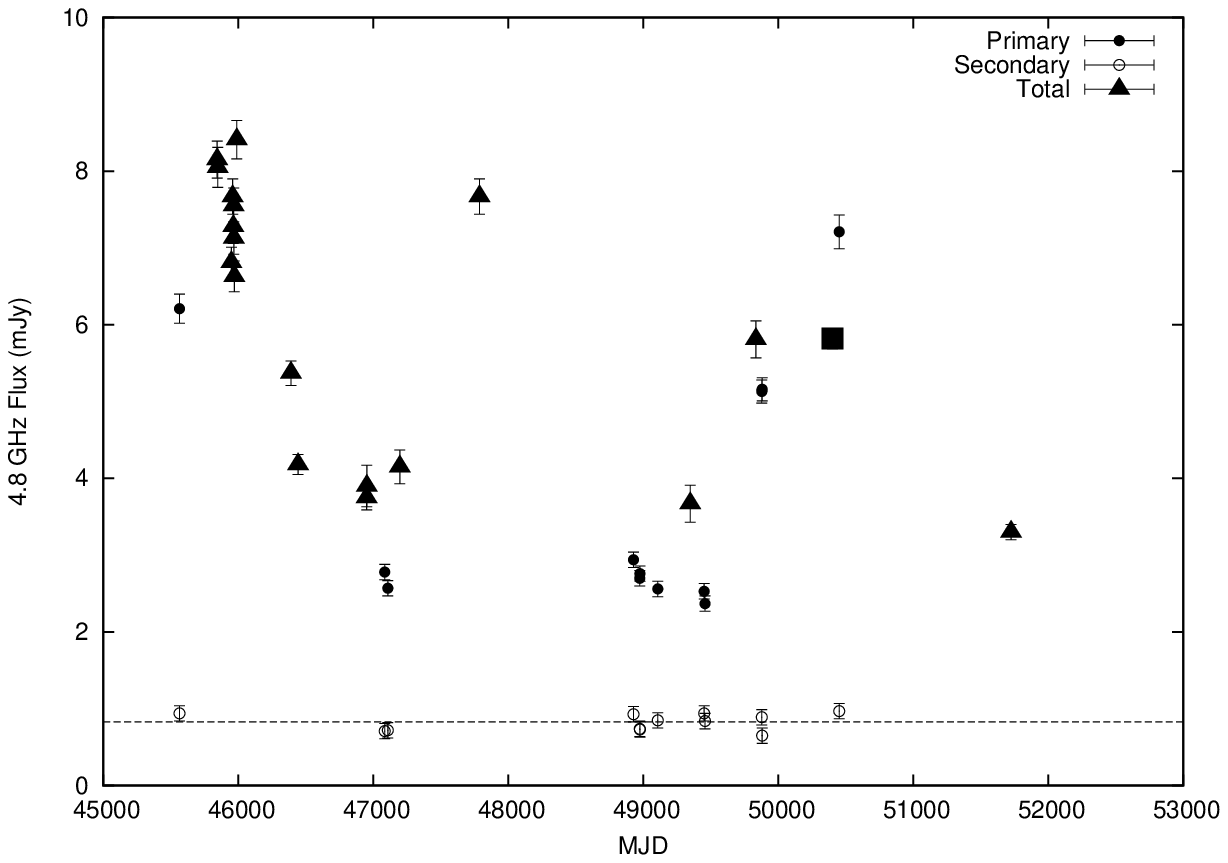}
\caption[]{The fluxes of \object{Cyg OB2 \#5} between 1983 and 2003 at
  8.4~GHz (top) and 4.8 GHz (bottom). Fluxes from the primary and NE
  are shown as solid and open symbols respectively. Those observations
  where the two components were {\em not} resolved as separate sources
  (i.e. primary$+$NE) are shown as triangles. The mean fluxes of NE at
  4.8 GHz ($0.83\pm0.11$~mJy) and 8.4 GHz ($0.50\pm0.12$~mJy) are
  shown (dashed line). The MERLIN observation at 4.8~GHz is shown as a
  square.
\label{fig:light_curves}}
\end{figure}

The fluxes of \object{Cyg OB2 \#5} at both 4.8~GHz and 8.4~GHz as a
function of time are shown in Fig.~\ref{fig:light_curves}. Through the
20 years of observations it is evident that the 4.8-GHz emission from
\object{Cyg OB2 \#5} has cycled through three cycles of high and low
emission with an approximate period of $\sim7$~yrs, as first noted by
\citet{Persi:1985}.  It is also clear that the radio emission from NE
shows no variation and the primary radio component is the source of
the variations in \object{Cyg OB2 \#5}.

At both 4.8~GHz and 8.4~GHz there were 12 and 11 epochs of VLA
observations respectively where NE was resolved separately from the
primary and a flux could be measured directly. These observations have
a weighted-mean flux at 4.8 GHz of $0.83\pm0.11$~mJy and
$0.50\pm0.12$~mJy at 8.4 GHz. The uncertainties are approximately
equal to our minimum image uncertainty of 0.1 mJy and all the fluxes
are within $1.7\sigma$ of these means. Hence, the 4.8~GHz and 8.4-GHz
fluxes from NE are taken to be constant.

In some observations the primary and NE are not resolved separately
and hence only a total flux for the system is determined. For these
observations, the primary source flux was determined, for the first
time, by subtracting the derived mean flux observed for NE from the
total observed flux at each frequency.

\begin{figure}[t]                            
\plotone{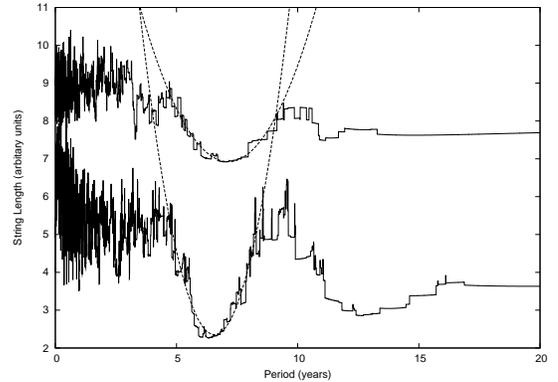}
\caption{\small String lengths for 17 8.4-GHz observations (top) and
  30 4.8-GHz observations (bottom). The derived period is $7.1\pm0.5$
  years for the 8.4 GHz observations and $6.6\pm0.3$ years for the 4.8
  GHz observations. Combined, these suggest a period of $6.7\pm0.3$
  years.}
\label{fig:stringlengths}
\end{figure}

Using fluxes from the primary at all observing epochs, the period of
variation was determined using a string-length technique
\citep{Dworetsky:1983}. This method was modified slightly to account
for relative uncertainties in the flux values \citep{vanloo:2008}. The
period and its uncertainty were estimated from the string-length data
as described by \citet{Fernie:1989}. The string lengths as a function
of period are presented in Fig.~\ref{fig:stringlengths} for both the
4.8-GHz and 8.4-GHz observations, along with the best-fit parabolas
from which the periods and associated uncertainties were derived. The
derived periods of the variations are $6.6\pm0.3$ years at 4.8 GHz and
$7.1\pm0.5$ years at 8.4 GHz. These are statistically consistent with
each other and averaging these two results leads to an estimated
emission variation period of $6.7\pm0.3$ years, which is remarkably
similar to the ``eyeball'' period derived in previous works.

The string-length technique was also used to search for periodicity in
the 4.8~GHz and 8.4~GHz observations consistent with the 6.6-day
orbital period of the binary. In this case no clear minimum was
discernible. This is not surprising as the sample rate of the
observations is much lower than the Nyquist frequency of a 6.6-day
period.

\subsection{Spectral Indices of the Emission}                  

\begin{figure}[t]                                                 
\plotone{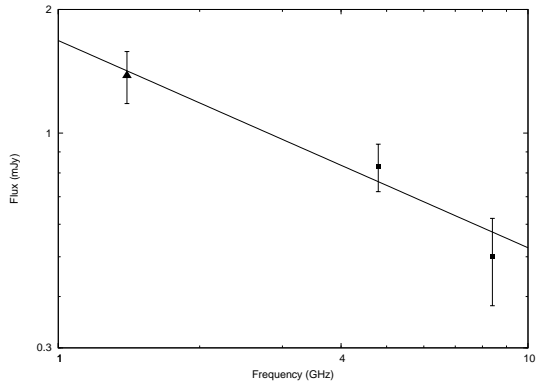}
\caption{\small The spectrum of the secondary source NE, based on the
  fluxes of $1.38\pm0.20$ mJy, $0.83\pm0.11$ mJy, and $0.50\pm0.12$
  mJy at 1.4 4.8 and 8.4 GHz respectively. The 1.4 GHz observation
  (triangle) is from MERLIN and the 4.8 and 8.4 GHz data (squares) are
  from the VLA. The best-fit spectral index is $-0.50\pm0.12$.}
\label{fig:sec_spectrum}
\end{figure}

The spectral index of a continuum spectrum is a characteristic of the
underlying radio emission. Assuming the flux $S_\nu$ at frequency
$\nu$ has a power-law behavior, $S_\nu\propto\nu^\alpha$, a weighted
least-square fit is used to determine the spectral index $\alpha$ of
each component in Cyg OB2 \#5.  MERLIN observations resolved the two
radio components separately, giving a 1.4-GHz flux for NE.  Since the
two 1.4-GHz fluxes are the same within uncertainties, and only $\sim1$
month apart, a mean flux of $1.38\pm0.20$ mJy was estimated for NE. In
combination with the mean fluxes determined at 4.8~GHz and 8.4~GHz
from the VLA observations, a spectral index of $-0.50\pm0.11$ was
derived for NE (Fig.~\ref{fig:sec_spectrum}). Higher frequency
detections of NE would be useful to demonstrate if the power-law
spectrum extends to higher frequencies.

\begin{figure}[t]
\plotone{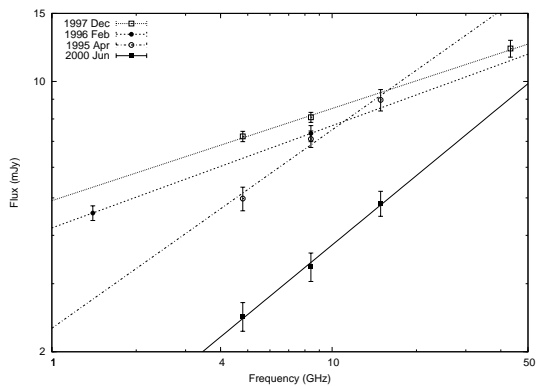}
\caption{\small The change in the continuum spectra of the primary
  radio component between the low-flux state (solid squares) toward
  the high-flux state (open squares), passing through intermediate
  states (open and solid circles).  The date of each set of
  observations is marked in the upper left corner.  The best-fit
  power-laws are shown for each set of data, with values of
  $0.60\pm0.04$ (solid line), $0.51\pm0.08$ (dot-dashed line),
  $0.26\pm0.04$ (dashed) and $0.24\pm0.01$ (dotted). The continuum
  spectrum during the low-flux state has a spectral index consistent
  with thermal emission from a stellar wind, whereas during the
  high-flux state the spectrum is flatter. It is argued in
  Sec.~\ref{sec:primary_emission} this is due to the addition of a
  non-thermal emission component to the thermal emission from the
  O-star binary.}
\label{fig:prim_spec}
\end{figure}

\begin{figure}[t]
\plotone{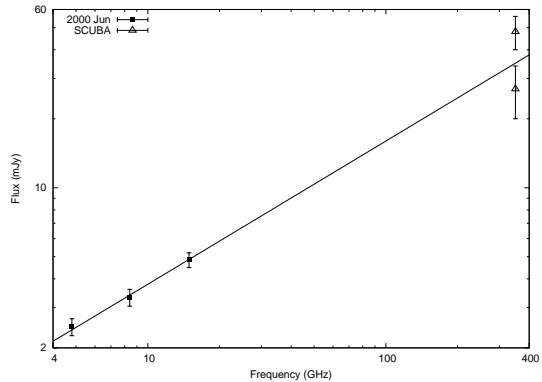}
\caption{\small The radio fluxes from the primary component during a
  low-emission state shown with the 350-GHz JCMT data. The best-fit
  spectrum has a slope $+0.63\pm0.04$, consistent with a stellar
  wind. The JCMT data were not obtained during the radio minimum 
  but fit this spectrum.}
\label{fig:prim_thermal}
\end{figure}

Using both the VLA and MERLIN observations it was possible to
determine the spectral index of the primary source at four different
epochs of the flux variation cycle. Using only one MERLIN observation
at 1.4 GHz and one VLA observation at 8.4 GHz from 1996 February
during a high emission state, a spectral index of $0.26\pm0.04$ was
calculated. Fluxes from the VLA at 4.8, 8.4 and 43.3 GHz during the
same high flux state around 1996 December 29 give a spectral index of
$0.24\pm0.01$. In a fashion, the spectral index during two different
low states was determined, with a spectral index of $0.51\pm0.08$ on
1995 April 27 and $0.60\pm0.04$ on 2000 June 30 from the VLA. The
15-GHz fluxes for the primary were determined by subtracting an
extrapolated 15-GHz flux for NE of $0.43$~mJy that assumes the
power-law spectrum for NE holds to 15 GHz. Examination of
Fig.~\ref{fig:sec_spectrum} suggests that a 15-GHz flux around 1~mJy
would present a unique spectrum not observed previously in a
WCR. Hence, it is estimated that the 15-GHz flux of NE, and hence the
primary component, has a systematic error of less than $\pm0.5$~mJy,
which still gives low-state continuum spectra consistent with thermal
wind emission.  These data and spectral indices are presented in
Fig.~\ref{fig:prim_spec}. The two 350-GHz data points are consistent
with the extrapolation of the radio data from the low-emission state
and a thermal stellar-wind spectrum (Fig.~\ref{fig:prim_thermal}).

\subsection{Proper Motion and Component Separation}
\label{sec:proper_motion}

A large number of observations spanning 20 years presents an
opportunity to check for evidence of proper motion of the radio
emission relative to the phase-reference source B2005+403.  There is
no evidence of proper motion in these data, with a scatter in
positions of 100 mas, consistent with the positional uncertainty
introduced by phase transfer at 4.8~GHz over $\sim5^\circ$ between the
phase calibrator and \object{Cyg OB2 \#5} using the VLA.

Absolute position uncertainties due to phase transfer do not impact
determination of the relative position of the sources in the
field. Using the models from seven epochs of 8.4-GHz
observations obtained in the VLA A-configuration, the position of the
NE relative to the primary was determined to be $+0.63\pm0.07\arcsec$
(East) and $+0.44\pm0.04\arcsec$ (North). This implies a component
separation of $0.77\pm0.08\arcsec$ at a position angle of
$55\pm2^{\circ}$ East of North, consistent with that derived from just
two observation epochs by \citet{Contreras:1997}.

\section{Discussion}
\subsection{The Primary and Variable Emission}
\label{sec:primary_emission}
The primary emission component is associated with the O-star binary
system and is the source of all the observed variations in the radio
emission. In the low state the primary radio emission is found to have
a spectral index of $0.60\pm0.04$ consistent with that expected for
thermal emission arising in a steady-state radially symmetric stellar
wind. The thermal emission must be reasonably constant in nature since
the 350-GHz observations were not both obtained during a radio minimum
yet are consistent with the stellar wind spectrum deduced from the
radio minimum observations, with a best-fit spectral index
$0.63\pm0.04$ across this broad frequency range
(Fig.~\ref{fig:prim_thermal}).

For a stellar wind, the mass-loss rate can be calculated from the
radio flux \citep[e.g.][]{Wright:1975}. Assuming the stellar wind has
a temperature of 10kK and a wind composition with ionic mean charge of
1, mean molecular mass of 1.5 and 1 electron/ion, and assuming all the
thermal emission arises from the binary stellar wind (see below for
discussion of another potential source of thermal emission), the
4.8-GHz flux at radio minimum of 2.5~mJy leads to a deduced mass-loss
rate of
\begin{equation}
\begin{split}
\dot{M}=& 3.4 \times 10^{-5} \times \nonumber\\
&\left( \dfrac{{\rm v}_{\infty}}{1500\,{\rm km\,s}^{-1}}
\right)
\left(\dfrac{d}{{1.7 \rm kpc}}\right)^{3/2}\dfrac{1}{\sqrt{F}}~~ 
M_\odot \, {\rm yr}^{-1}, \nonumber
\end{split}
\end{equation}
where v$_{\infty}$ is the terminal wind velocity, $F$ is the volume
filling factor \citep[e.g.][]{Abbott:1981}, and $d$ is the distance to
the source. The terminal velocity of 1500 km\,s$^{-1}$ is adopted from
\citet{Conti:1999} based on the absorption component of the P-Cygni
profile of the He\,{\sc i} 1.083~$\mu$m line. It should be noted that
the emission component of that line is very strong, and broader than
those of other supergiants observed by them or in the atlas of
\citet{Groh:2007}, more closely resembling line profiles in WN-type
spectra. 

The mass-loss rate derived here is consistent with other values
determined for \object{Cyg OB2 \#5}:
$3.3\times10^{-5}$~M$_\odot$~yr$^{-1}$ from UV lines
\citep{Howarth:1989}, $2.5\times10^{-5}$~M$_\odot$~yr$^{-1}$ from
infrared observations \citep{Persi:1990},
$3.7\pm1.3\times10^{-5}$~M$_\odot$~yr$^{-1}$ based on the 43-GHz flux
\citep{Contreras:1996}, adjusted for the terminal velocity of
\citet{Conti:1999} and for slight differences in adopted distance.
More recently the reliability of these methods has been called into
question as they tend to overestimate the actual mass-loss rate if the
wind is clumpy \citep[e.g.][]{Mokiem:2007,Fullerton:2006}. The most
recent estimate of mass-loss rate in \object{Cyg OB2 \#5} is from
\cite{Linder:2009}, who suggest a rate of
$2.1\pm0.6\times10^{-5}$~M$_\odot$~yr$^{-1}$ for the binary system
based on the observed rate of period change. This is independent of
distance and wind clumping but there are other processes which may
affect the period. Nevertheless, all the mass-loss rate estimates are
consistent with each other, though notably higher than values derived
from model atmosphere fits for stars of similar spectral type
\citep{Mokiem:2007}, yet comparable to those of WR stars
\citep{Crowther:2007}. This is consistent with the unusual strength of
the He\,{\sc i} 1.083-$\mu$m line in \object{Cyg OB2 \#5}, indicating
that the system, or one of its components, has a fast, heavy stellar
wind comparable to those observed in WR stars.

As the flux increases from the low state toward the high state the
continuum spectrum flattens out (see Fig.~\ref{fig:prim_spec}). This
is consistent with the findings of \citet{Persi:1990}, who attributed
the flattening to non-thermal emission from an expanding plasmon
associated with the binary system. The model they produced resulted in
broad agreement with the general variations in the data, though the
fit to their data was poor.

Most recently, 3D-hydrodynamical models of O-star binary systems with
periods of a few days have demonstrated that radio flux variations and
the flatter spectral index can result from variable {\em thermal}
emission arising in a WCR between the binary stars
\citep{Pittard:2009}. \object{Cyg OB2 \#5} is a contact system
\citep{Leung:1978} and the nature of a wind-collision region in such a
system is unclear. However, such a region will emit thermal emission
that is likely variable. Whether the flux and variation amplitudes can
be attained in such a region is discussed further in
Sec.~\ref{sec:thirdstar},

An alternative model for the primary radio emission component is
proposed here, where the lower spectral index during high emission is
the result of the addition of a non-thermal component to the thermal
emission from the binary system giving a ``composite'' spectrum. Such
a model has been successfully applied to describe the relatively flat
continuum spectra of some Wolf-Rayet stars
\citep[e.g.][]{Chapman:1999} where the non-thermal emission
arises in a WCR between the wind of the WR star and that of a massive
companion star.

For a system consisting of a non-thermal source embedded in a stellar
wind plasma, the total observed flux is given as a function of
frequency $\nu$ and at epoch $t$ by

\begin{displaymath}
S_{obs}(\nu,t) = S_{th}(\nu) + S_{nt}(\nu,t)~~~~~{\rm mJy}.
\label{eqn:obs_flux}
\end{displaymath}
It is assumed the constant thermal emission component, $S_{th}(\nu)$,
has spectral index of $+0.6$ and a flux at 4.8~GHz of 2.5~mJy, deduced
from the the primary source during the low emission state. Hence
\begin{displaymath}
S_{th}(\nu) = 2.5 \left(\dfrac{\nu}{4.8}\right)^{0.6}.
\label{eqn:th_flux}
\end{displaymath}
The non-thermal emission component of the total flux, $S_{nt}(\nu,t)$,
is modelled as
\begin{displaymath}
S_{nt}(\nu,t) = S_{4.8}(t) \left(\dfrac{\nu}{4.8}\right)^{\alpha} e^{-\tau(\nu,t)},
\label{eqn:nonth_flux}
\end{displaymath}
where $S_{4.8}(t)$ is the intrinsic 4.8-GHz flux of the non-thermal
source at epoch $t$, $\alpha$ is the spectral index of the non-thermal
emission assumed to be constant, and $\tau(\nu,t)$ is the
line-of-sight free-free opacity through the stellar wind to the
non-thermal source at frequency $\nu$ and epoch $t$, approximated by
\begin{displaymath}
\tau(\nu,t) \approx \tau_{4.8}(t)\left(\dfrac{\nu}{4.8}\right)^{-2.1}
\end{displaymath}
where $\tau_{4.8}(t)$ is the 4.8-GHz line-of-sight free-free opacity
at epoch $t$.

The line-of-sight opacity is dependent on the geometry of the
line-of-sight to the non-thermal emission. Here, the case of a
non-thermal source in orbit about the binary is
considered. \citet{Williams:1990} derived the varying free-free
opacity along a line-of-sight to a non-thermal source orbiting in the
circumbinary wind of the massive WR+O binary WR\,140. Following
Eqns.~12 and 14 in \cite{Williams:1990}, the opacity is dependent on
the orbit inclination ($i$), argument of periastron ($\omega$), as
well as the epoch-dependent true anomaly ($f$) and the separation of
the orbiting source from the companion star ($r$) in units of
semi-major axis distance ($a$) such that
\begin{equation}
\label{eqn:williams}
\begin{split}
\tau_{4.8}(t)=&\dfrac{ \xi \sec i}{ 2\Delta r^{3}\cos^{3}(\omega +f)} \times  \\
              & \biggl(\sin (\omega +f)\cos (\omega +f)\tan i  + \\ 
              & (1+\tan^{2}i)\arctan \biggl( \dfrac{-\sqrt{\Delta }}{\tan (\omega +f)\tan i}\biggr)\biggr), 
\end{split}
\end{equation}
where
\begin{displaymath}
\Delta =1+\tan ^{2}(\omega +f)+\tan ^{2}(i)
\end{displaymath}
and $\xi$ is a constant proportional to the square of the ion
density in the stellar wind at a radius equal to the semi-major axis
$a$, and related to $C_{\rm ff}$ in \cite{Williams:1990} by
\begin{displaymath}
C_{\rm ff}=\xi \left(\dfrac{\nu}{4.8}\right)^{-2.1}.
\end{displaymath}

The intrinsic non-thermal flux $S_{4.8}(t)$ is expected to depend on
the local conditions e.g. electron density, which will vary as the
source moves through the dense circumbinary wind. This may be
approximated by assuming a simple power-law relation with separation,
namely
\begin{displaymath}
S_{4.8}(t) = S'_{4.8} r^{-s},
\label{eqn:s48propto}
\end{displaymath}
where $S'_{4.8}$ is the non-thermal flux when the separation is equal
to $a$, and $s$ is the power-law index.  These definitions, along with
the analytic solution to equation \ref{eqn:williams}
\citep[cf.][]{Williams:1990}, allow $S_{obs}(\nu,t)$ to be determined
as a function of the orbital phase of the non-thermal source orbiting
the binary system.

\begin{figure}[t]
\plotone{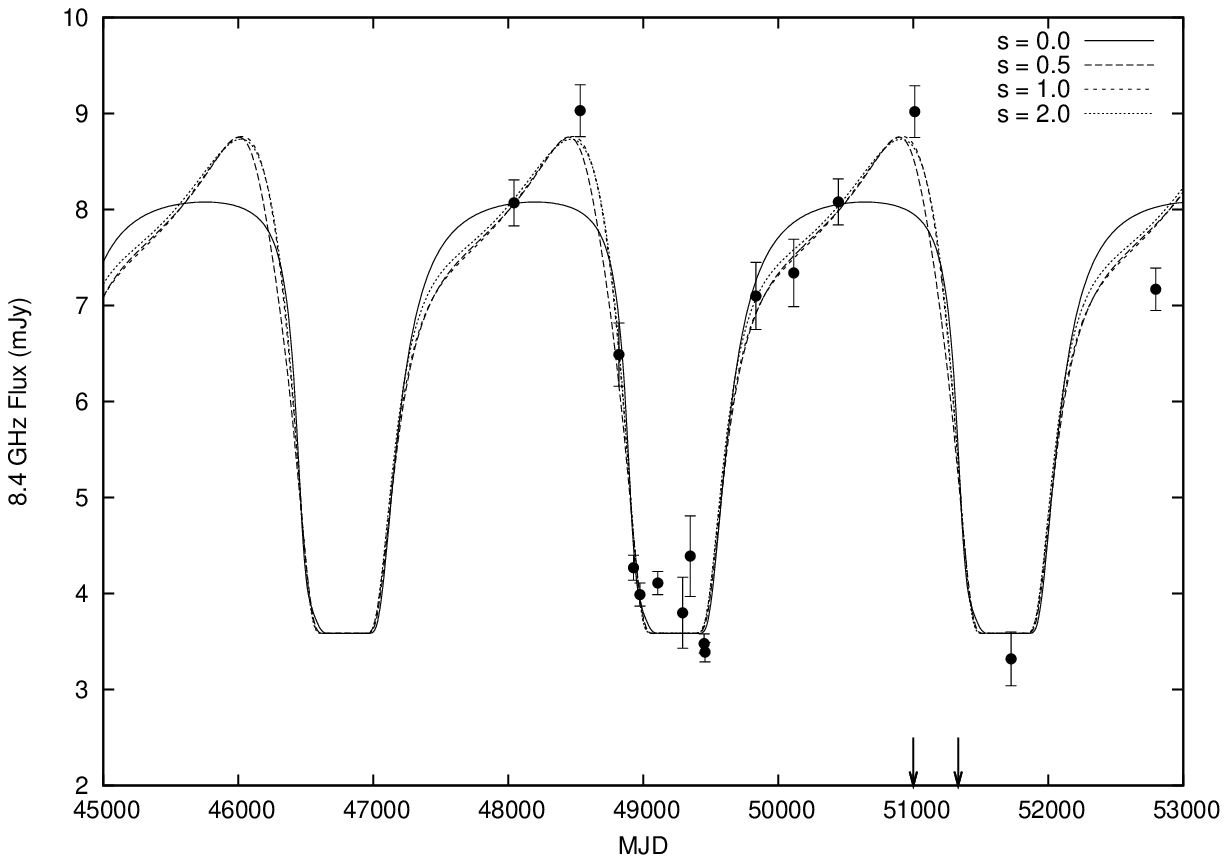}
\plotone{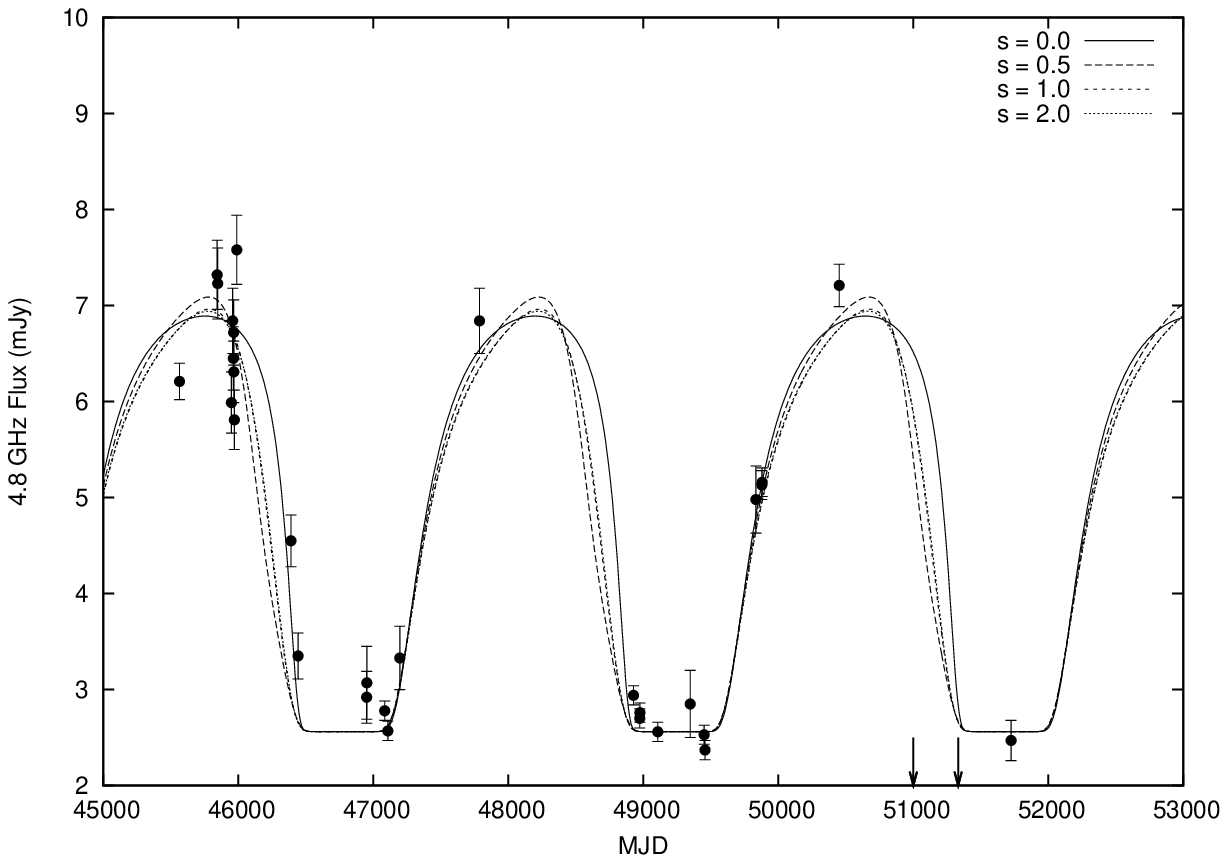}
\caption{\small The best-fit orbiting non-thermal source model is shown
  against the observed fluxes of the primary at 8.4 GHz (top) and 4.8
  GHz (bottom) for the $s=0, 0.5, 1$, and 2 models.}
\label{fig:spectrum_model} 
\end{figure}

\begin{figure}[t]
\plotone{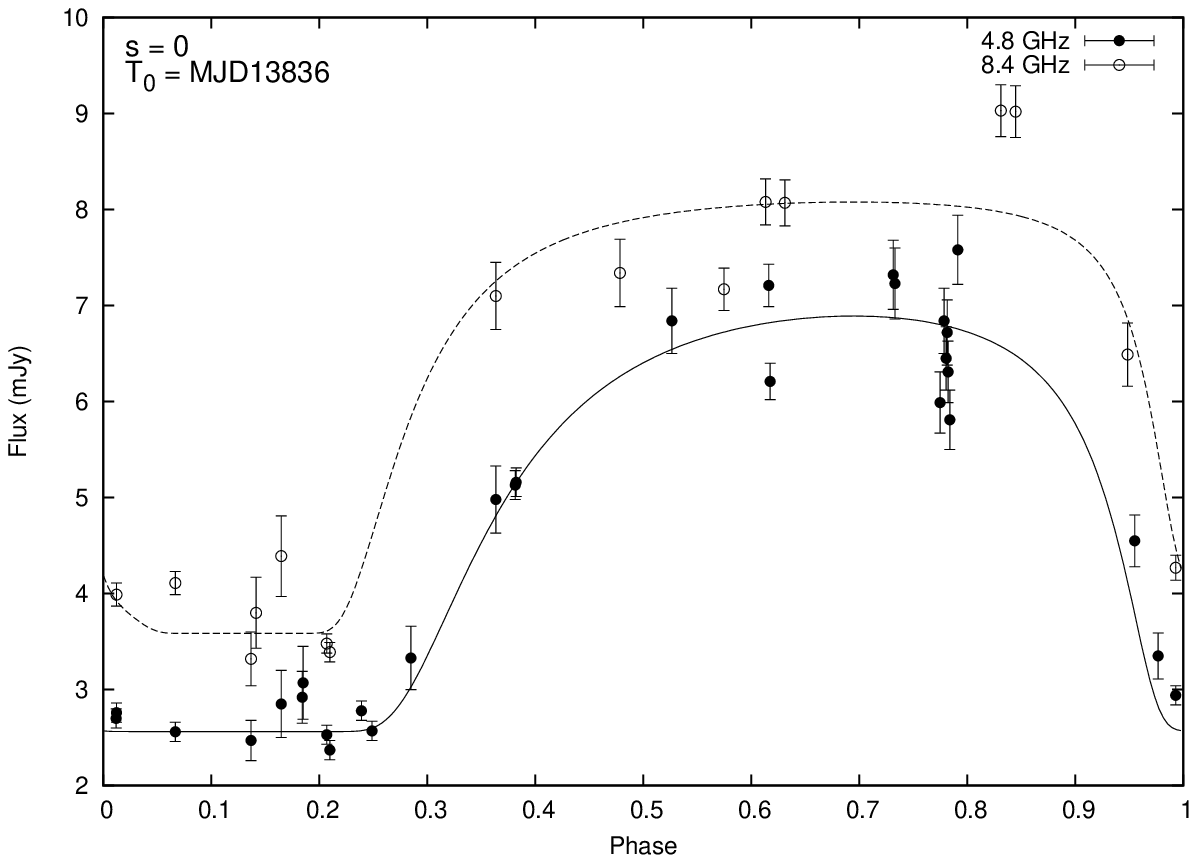}
\plotone{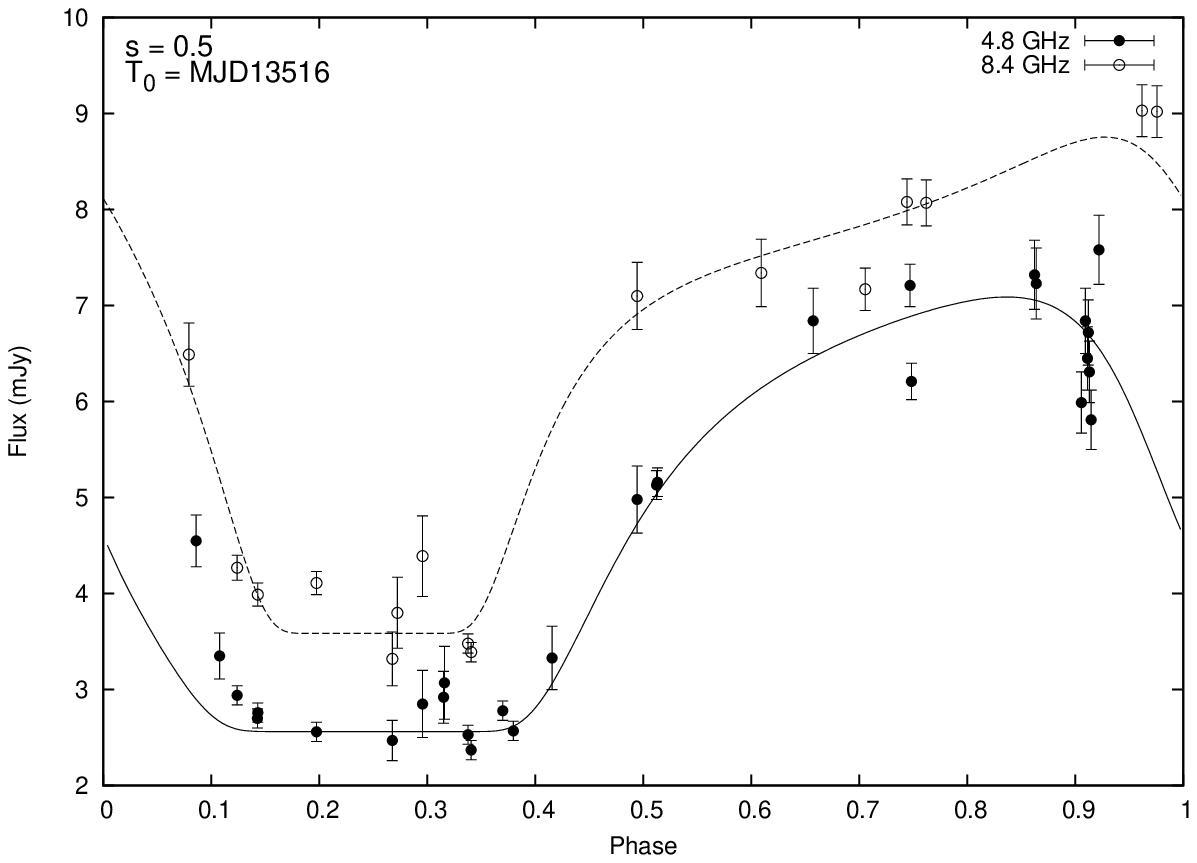}
\caption{\small The best-fit orbiting non-thermal source model is
  shown for the $s=0$ (top) and $s=0.5$ (bottom) cases at 4.8 GHz
  (solid line) and 8.4 GHz (dashed line) against the fluxes of the
  primary, phased with a period of 6.7-year. Parameters for each model
  are given in Table~\ref{tab:bestmodel_parms}.}
\label{fig:phase_model}
\end{figure}

A standard Levenberg-Marquart $\chi^{2}$-minimization technique was
applied to both the 4.8-GHz and 8.4-GHz fluxes of the primary
component to determine values for the seven free parameters in the
model for each of the cases $s=0,0.5,1,$ and $2$, assuming an orbital
period of 6.7 years. \cite{Dougherty:2003} suggest $s=0.5$ for the
non-thermal luminosity of a WCR. The resulting model parameters are
given in Table~\ref{tab:bestmodel_parms} and the light curves arising
from these models are plotted in Fig.~\ref{fig:spectrum_model}.
Fig.~\ref{fig:phase_model} shows the models for the $s=0$ and $s=0.5$
cases folded into the 6.7-year period.

\begin{deluxetable*}{cccccccc}
\tablewidth{0pt}
\tabletypesize{\scriptsize}
\tablecaption{Best model-fit parameters for the orbiting non-thermal source
  model.\label{tab:bestmodel_parms}}
\tablehead{\colhead{$s$}&\colhead{$S'_{4.8}$}&\colhead{$\alpha$}&\colhead{$\omega$}&\colhead{$i$}&\colhead{$e$}&\colhead{$T_{0}$}&\colhead{$\xi$}\\
\colhead{}&\colhead{(mJy $a^{s}$)}&\colhead{}&\colhead{$(^{\circ})$}&\colhead{$(^{\circ})$}&\colhead{}&\colhead{(MJD)}&\colhead{}}
\startdata
\tableline
\tableline
0  & $5.3\pm0.5$ & $-0.18\pm0.25$ & $319\pm3$ & $90\pm40$ & $0.69\pm0.04$ & $53836\pm35$ & $0.48\pm0.13$\\
0.5 & $6.4\pm0.6$ & $-0.34\pm0.26$ & $315\pm3$ & $90\pm25$ & $0.44\pm0.04$ & $53516\pm34$ & $0.62\pm0.12$\\
1 & $7.1\pm0.6$ & $-0.42\pm0.28$ & $352\pm5$ & $88\pm46$ & $0.23\pm0.04$ & $53636\pm30$ & $1.36\pm0.22$\\
2 & $7.6\pm0.7$ & $-0.47\pm0.30$ & $23\pm5$ & $85\pm48$ & $0.11\pm0.05$ & $53737\pm31$ & $1.64\pm0.25$\\
\enddata
\end{deluxetable*}

Each of these simple models show a good fit to the data and are
effective in matching observations at both frequencies across all four
observed emission cycles through the 20 years of observation.  The
reduced-$\chi^{2}$ values range from 4.4 to 4.6 indicating that each
model has its flaws, with the $s=0$ case being formally the
best-fit. The $s=0$ case corresponds to intrinsic non-thermal
emission that is constant throughout the orbit i.e. the variation is
caused entirely by the varying line-of-sight opacity. In this case the
best-fit model is a good fit to the observations except it is too low
to match the peak emissions at 8.4 GHz. Introduction of non-zero
values for $s$ gives models able to match these sharp emission peaks
but the resulting model flux is too low to match the 4.8 GHz
observations when the primary flux is in decline at MJD~46309 and
MJD~46443 from the high to low emission state, just prior to radio
minimum (around deduced orbital phase $\sim0.1$). 

\citet{Williams:1990} could not fit the radio flux variations in
WR\,140 with this model and a single value of $C_{\rm ff}$
(corresponding to our $\xi$). This was attributed to the very
different densities of the WR and O-star winds traversed by the
line-of-sight at different orbital phases. The quality of the fits to
the \object{Cyg OB2 \#5} radio-flux variations with a single value of
$\xi$ suggests that the stellar winds in this system have comparable
densities, consistent with stars having comparable mass-loss rates
rather than the $\sim30$-fold ratio between the WR and O-star winds in
WR\,140.

In all four models the stellar wind is completely opaque during the low
emission state. Unfortunately the radio observation at both
frequencies are not evenly distributed across the 20 years of
observations and there are large gaps in the phase coverage of the
observed light curves. Consequently the model fitting is unevenly
focused by the group of observations obtained during the low emission
states. Based on the $s=0$ model, the last low emission state occurred
around 2007 February, and the next high emission state will occur around
2010 November.

The inclination across all models is consistent with an eclipsing
orbit, though this parameter is poorly constrained.  Likewise, even
though the non-thermal spectral index tends to become more negative
with increasing $s$, it is also poorly constrained. The non-thermal
flux $S'_{4.8}$ increases with increasing $s$ as would be expected: a
greater flux fall-off with separation would require stronger emission
to match the observations. Varying $s$ ranges the time of periastron
passage ($T_{0}$) by up to a year between the $s=0$ and $s=0.5$ cases,
as evident in Fig.~\ref{fig:phase_model}. The remaining $T_{0}$  
values fall within this range. By far the greatest impact of varying
the $s$ parameter is on the deduced orbit eccentricity. For $s=0$, the
orbit eccentricity is high with $e\sim0.7$, and as $s$ increases the
eccentricity decreases sharply, with $e\sim0.1$ for the $s=2$ case. As
noted above, many of the observations were made during the low-state
when the stellar-wind is opaque and the non-thermal emission does not
contribute to the observed flux being fit by the model. This
contributes to the uncertainties in the fitted parameters.

\subsection{Evidence for a Third Star?}
\label{sec:thirdstar}
A non-thermal source orbiting the binary system requires a star 
(hereafter Star C) to be in a 6.7-year orbit around the binary. 
This star could contribute the non-thermal radio emission via a WCR
arising from the collision of its own stellar wind with the wind from
the O+O star binary \citep[e.g.][]{Eichler:1993}. Such WCRs have been
observed directly in some WR+O star and O+O star binary systems
\citep[e.g.][and references therein]{Dougherty:2006}. 
Alternatively, the non-thermal emission may arise from the putative
third star directly, e.g. a compact object.

Given the high luminosities of the two supergiants in the binary and
emission from circumstellar material, it will be very hard to detect
the proposed third star directly, let alone measure its orbit.
Instead, the radial velocities (RVs) of the central binary are
examined to search for reflex motion due to it being in an orbit
with Star C, as suggested by the radio observations.

The RVs measured by \citet{Rauw:1999} come from four observing 
runs, each between 1 and 4 weeks duration and separated by about a
year. As the variations should coincide with a period near 6.7 years,
each of these runs is treated as a single observation. A fifth
observation comes from the first five RVs measured by
\citet{Bohannan:1976} in the space of a month 23 years earlier.  For
each RV observed from the primary\footnote{The secondary has not been
used as \citet{Rauw:1999} find a significantly lower $\gamma$-velocity
for it in their orbit solution and attribute this to formation of the
absorption lines in the wind}, the residual (O--C) was calculated from
the orbit by \citet{Rauw:1999} (based on all the RVs) and formed the
average (O--C) for each run. These are given in Table \ref{tab:OmC},
together with the radio orbital phases calculated using P = 6.7 years
and $T_{0}$ for the $s=0$ model from Table \ref{tab:bestmodel_parms}.

\begin{deluxetable}{ccccc}
\tablewidth{0pt}
\tabletypesize{\scriptsize}
\tablecaption{Mean RV deviations (O--C) from the O+O orbit as a 
         function of radio phase for five observing runs.\label{tab:OmC}}
\tablehead{
\colhead{MJD}&\colhead{$\phi$}&\colhead{n (RVs)}&\colhead{mean (O--C)}&\colhead{$\sigma$}
}
\startdata
41150    &  0.81  &   5   &  14.9  &    9.0  \\
49567    &  0.26  &   11  &  -7.6  &    6.7  \\
49914    &  0.40  &   7   &  -2.2  &    6.6  \\
50316    &  0.56  &   3   &  -3.7  &    7.3  \\
50640    &  0.69  &   4   &  8.6   &    8.7  \\
\enddata
\end{deluxetable}

A systematic increase of RV between phases 0.26 and 0.89 is seen,
implying that the O+O binary moves away from us more rapidly.  This
implies that Star C moves towards us more rapidly in this phase 
interval so that the circumbinary extinction to the non-thermal
radio source diminishes, consistent with it brightening during this
orbital phase.

\begin{figure}[t]
\plotone{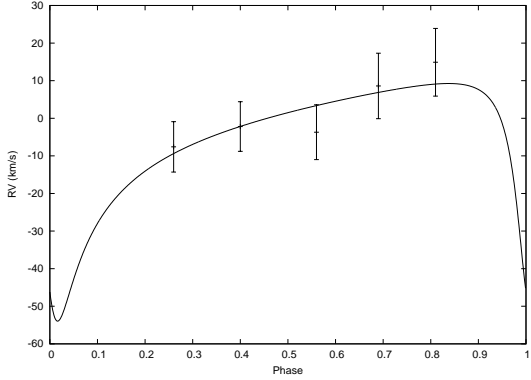}
\caption{\small Comparison of observed (O--C) residuals, plotted with
  2-$\sigma$ error bars against radio phase, and the RV curve for the
  reflex motion corresponding to the elements of the $s=0$ model
  (i.e. with $\omega$ shifted by 180$\degr$ to reflect the location of
  the binary system in the orbit, rather than the 3rd star), with
  $K=32\pm17$~km\,s$^{-1}$ and $\gamma=-5.9\pm4.7$~km\,s$^{-1}$
  giving the best-fit of the observables.}
\label{fig:RVOmC}
\end{figure}

The run of mean (O--C) with phase is compared with the reflex motion
of the O+O binary in orbit with Star C following the orbital elements
of the embedded non-thermal radio source from the $s=0$ case (see
Fig.~\ref{fig:RVOmC}). Fitting
\begin{displaymath}
{\rm v}_{r} = \gamma+K_{{\rm O+O}}\bigl( e \cos\omega + \cos(f + \omega) \bigr)
\end{displaymath}
for $K_{{\rm O+O}}$ and systemic velocity $\gamma$, gives $K_{{\rm
O+O}} = 32\pm17$~km\,s$^{-1}$ and $\gamma=-5.9\pm4.7$~km\,s$^{-1}$,
both very uncertain given the uncertainties in the (O--C)s and the
shape of the RV curve in the phase range of the observations. The
non-zero $\gamma$ is a consequence of not weighting the mean (O--C)s
by the rather unequal numbers of observations from which they were
deduced. The mass function $f(m)$ can be derived from $P$ (in
days) and $K$ (in km\,s$^{-1}$) from
\begin{equation} 
\begin{split}
f(m) = & \dfrac{m^3_{\rm C}\sin^3(i)}{(m_{{\rm O+O}}+m_{\rm C})^2} \\
     = & 1.036 \times 10^{-7} \left(1 - e^{2}\right)^{3/2} K^{3} P, \nonumber
\label{eqn:massfn}
\end{split}
\end{equation}
allowing an estimate of the mass, $m_{\rm C}$, of Star C. From the
data here, $f(m)=3.2^{+8.2}_{-2.8}~{\rm M}_{\odot}$. Assuming
$\sin(i)=1$ and adopting $m_{{\rm O+O}} = 41.5\pm3.4$~M$_{\odot}$ from
\citet{Linder:2009}, this gives $m_{\rm C} =
23^{+22}_{-14}$~M$_{\odot}$ for Star C. The large uncertainty in the
mass stems from the high relative uncertainty in $K$ and the
$K^3$-dependence of $f(m)$.

With a paucity of observations, most especially at phases of the
putative orbit where the radial velocity changes most dramatically,
the uncertainties in this analysis are high. However, until further
observations can be obtained to test this analysis, it remains a
tantalizing piece of evidence that \object{Cyg OB2 \#5} is a tertiary
system rather than a binary.

Additional support for Star C comes from X-ray data that reveal a hard
component. \cite{Linder:2009} suggest this is likely to arise in a WCR
and argue that in such a compact binary like \object{Cyg OB2 \#5} the
stellar winds of each component are far from achieving terminal
velocity and the resulting X-ray emission would be rather soft. Hence
they suggest a WCR would be between the binary and another star.  Such
a WCR between the binary and Star C provides a ready source for the
non-thermal radio emission in the model described in
Sec.~\ref{sec:primary_emission}. Such a mechanism has been widely
established for WR stars and many O star systems that exhibit
non-thermal emission \citep{deBecker:2007}.  A mass of
$\sim23$~M$_\odot$ for Star C is consistent with a late O/early B-type
star, which would have a sufficiently strong stellar wind to produce a
WCR with the binary system wind in an orbit of size $\sim14$~AU.

The recent models of thermal emission in WCRs by \cite{Pittard:2009}
raise the possibility that variable plasma density in a WCR between
the binary and star C might account for the variable radio emission.
Given the separation of the 6.7-year orbit, a WCR between the binary
and Star C is undoubtedly adiabatic (certainly away from orbit
periastron) and thus any thermal free-free emission in the WCR will be
optically thin, with a spectral index of $-0.1$. In combination with
the stellar wind continuum from the binary system and Star C,
this could result in a continuum spectrum that is flatter than a
stellar wind spectrum if the thermal flux from the WCR is sufficiently
high. The fluxes in the simulations of \cite{Pittard:2009} are two to
three orders of magnitude less than observed in \object{Cyg OB2 \#5},
attributable to the use of mass-loss rates $\sim$two orders of
magnitude less than deduced here for Cyg OB2 \#5. The optically-thin
thermal flux scales as the total number of ions in the WCR, which is
$\propto \dot M^{2}/D$, where $D$ is the distance from the binary to
the WCR. Since the deduced mass-loss rate and separation in
\object{Cyg OB2 \#5} are respectively $\sim100$ and $\sim40$ times
those used in the simulations \citep[cf. model cwb2
in][]{Pittard:2009}, the optically-thin flux from the WCR could be a
similar order of magnitude as the observed fluxes. However, in this
adiabatic scenario the flux variations scale as $D^{-1}$, and a
highly eccentric orbit, as deduced in the $s=0$ model presented above,
spends the bulk of the orbit near apastron where $D$ changes
little. Hence, the flux would change little, contrary to the
observations. Thus variable thermal emission from the WCR alone can
not account for the radio light curves in \object{Cyg OB2 \#5}, though
free-free opacity through the circumbinary wind undoubtedly plays a
role. This possibility needs to be explored further, though the types
of models described by \cite{Pittard:2009} are beyond the scope of
this paper.

Lastly, an alternative source of the X-ray and non-thermal
radio emission could be a compact object such as a neutron star,
though the estimated mass from the reflex motion analysis implies this
possibility is remote. Though it is not clear how a compact object
produces the radio emission, stellar wind accretion onto a
$\sim2$~M$_\odot$ compact star could generate the observed X-ray
luminosity if gravitational potential can be converted to X-ray power
efficiently. The gravitational capture radius of a neutron star of
mass $M_n$ for a stellar wind of velocity v$_\infty=
1500$~km\,s$^{-1}$ is given by
\begin{displaymath} 
R_g = \dfrac{2GM_n}{{\rm v}_\infty^2}\sim 2.3\times 10^{10}~{\rm cm}.  
\end{displaymath} 
Assuming the gravitational energy of the captured stellar wind is
converted to X-ray emission with efficiency $\epsilon$, then luminosity
would be
\begin{displaymath}
L_x=\dfrac{GM_n \pi R_g^2 F_m(r)\epsilon}{R_n} \sim 3.1 
\times 10^{41} F_m(r)\epsilon~~{\rm erg~s}^{-1}, 
\nonumber
\end{displaymath} 
where $F_m(r)$ is the mass flux of stellar wind at distance $r$. For
an orbit major axis of 14~AU and an eccentricity of 0.7 ($s=0$ model),
periastron separation is 4.2~AU at which distance
$F_m=4.3\times10^{-8}$~g\,cm$^{-2}$\,s$^{-1}$ for a stellar wind
mass-loss rate of $3.4 \times 10^{-5}$~M$_\odot$~yr$^{-1}$. This gives
$L_x \sim 3.4$~L$_\odot$ for $\epsilon=1$, consistent with the
$L_x=1.5$~L$_\odot$ derived by \cite{Linder:2009} and adjusted to a
distance of 1.7~kpc.

\subsection{The Secondary Source, NE}
The spectral index of $-0.50\pm0.11$ derived here indicates the radio
emission from NE is non-thermal. This was previously suggested by
\citet{Contreras:1996} from a limit to the spectral index of
$-2.4\pm0.6$ deduced from one epoch of observations at 4.8 and 8.4 GHz
in 1994 April. \citet{Contreras:1997} were the first to note that NE
lies directly between the \object{Cyg OB2 \#5} binary and a B-type
star (Star D), $\sim0.9\arcsec$ to the NE. This led them to propose
that NE is the result of a WCR between the stellar wind of the
\object{Cyg OB2 \#5} binary system with that from Star D. They further
argued that the separation of NE from both the primary and Star D was
consistent with the expected relative wind momenta of the binary and
the B-type star. This assertion is re-examined here based on seven
epochs of 8.4-GHz observations of the primary and NE.

The position of the WCR relative to the positions of the sources of
the colliding winds is given by
\begin{displaymath}
r_{\rm O+O} = \left ( 1 - \dfrac{ \eta^{1/2}}{1 + \eta^{1/2}}\right ) D
\end{displaymath}
\citep[e.g.][]{Eichler:1993} where $D$ is the separation between the
primary O-star binary (or triple) system and the B star, $r_{\rm O+O}$
is the distance from the WCR to the primary system and $\eta$ is the
the wind-momentum ratio of the two stellar winds given by
\begin{displaymath}
\eta = \dfrac{\dot{M}_{\rm D} {\rm v}_{\rm D}} {\dot{M}_{\rm O+O} {\rm v}_{\rm O+O}}.
\end{displaymath}
Here $\dot{M}_{\rm O+O}$, $\dot{M}_{\rm D}$, ${\rm v}_{\rm O+O}$ and
${\rm v}_{\rm D}$ are the mass-loss rates and terminal-wind velocities
of the O-star binary and Star D respectively. The separation of the
O-star binary and the B-type companion is $D=0.93\pm0.02\arcsec$ as
determined from a weighted average of separations deduced from optical
and IR observations (Sec.~\ref{sec:irobs}).  Combined with $r_{\rm
O+O}=0.77\pm0.08\arcsec$ measured here (Sec.~\ref{sec:proper_motion}),
$\eta = 0.04^{+0.08}_{-0.03}$, where a Monte-Carlo method was used to
determine the uncertainty in $\eta$ as it is an ill-behaved function.
Adopting $\dot{M}_{\rm O+O} = 3.4 \times 10^{-5}$~M$_{\sun}$ yr$^{-1}$
with v$_{\rm O+O} = 1500$~km\,s$^{-1}$ \citep{Conti:1999} leads to
\begin{displaymath}
\dot{M}_{\rm D} = 0.5{\rm-}6.1 \times 10^{-6} \left( \dfrac{1000\,{\rm
    km}\,{\rm s}^{-1}}{{\rm v}_{D}} \right) {\rm M}_{\sun} \, {\rm
    yr}^{-1}. 
\end{displaymath}
Keeping in mind the mass-loss rate of the binary is high compared with
values deduced from model atmospheres, and the wind terminal speed for
Star D is unknown, it is noted that only at the lower extremum is the
mass-loss rate consistent with those expected for late-O/early-B
supergiants, with early-B dwarfs having mass-loss rates around an
order of magnitude lower \citep{Mokiem:2007}, though it is noted that
the IR photometry of Star D suggest a star that is more luminous than
anticipated for a B0 dwarf. The need to consider the extremum
mass-loss rate in order to account for the relative location of the
binary, NE and Star D raises the question of whether NE is truly a
WCR. However, the Wolf-Rayet system WR\,147 provides a ready example
of an early B-type dwarf star providing a sufficiently dense wind to
give a readily observed WCR with the dense wind of a WN8 companion
\citep{Williams:1997}.

The proximity (in space) of Star D and the \object{Cyg OB2 \#5} binary
can be tested by comparing their respective reddenings. Combining the
$K$ magnitude determined in Sec.~\ref{sec:irobs} with the visual
magnitude $V=13.1\pm0.4$, derived by \citet{Contreras:1997}, gives
$(V-K) = 5.5\pm0.4$ for Star D, assuming it does not vary.  This
implies a reddening of $A_V \simeq 6.9\pm0.5$ for Star D, greater than
that ($A_V = 5.7\pm0.3$) implied by $(B-V) = 1.6\pm0.1$ measured by
\cite{Contreras:1997}. In spite of this apparent discrepancy between
visual and visual/IR-determined reddenings, the fact that the
reddening estimates bracket that of \object{Cyg OB2 \#5} ($A_V =
6.4$), there is no reason to rule out an association between the
binary and Star D, and hence the possibility that \object{Cyg OB2 \#5}
is a quadruple system.

It is possible to estimate the luminosity of a WCR based upon the
kinetic energy of the two colliding winds. The surface area of the WCR
can be approximated as a spherical cap of diameter $\pi r_{\rm D}$
\citep{Eichler:1993} where $r_{\rm D} = D - r_{\rm O+O}$ is the
distance from Star D to the WCR. Using this area combined with the
kinetic luminosity of the stellar wind of the binary, $L_{\rm O+O} =
2.4\times10^{37}$~erg\,s$^{-1}$, leads to luminosity $L_{WCR} = 6.4
\times 10^{35}$~erg\,s$^{-1}$ entering the WCR. The radio synchrotron
luminosity $L_{syn}$ arising from the WCR, is estimated from
$L_{syn}\sim10^{-8} L_{WCR}$ \citep{Chen:1994, Pittard:2006}, giving
an estimated synchrotron luminosity from the WCR of
$\sim6\times10^{27}$ erg\,s$^{-1}$.  The radio luminosity of NE is
estimated by integrating the observed continuum spectrum. Assuming a
power-law spectrum between 0.1 - 100 GHz gives a synchrotron
luminosity of $2.5\times10^{26}$~erg\,s$^{-1}$ for a distance of
1.7~kpc. Considering this is an order-of-magnitude argument, the
synchrotron luminosity of the NE is closely consistent with that
anticipated from a WCR.
 
An alternative model for NE is that of a background radio source in
chance alignment with \object{Cyg OB2 \#5} and Star D. It is difficult
to refute this possibility unequivocally, though the probability of
such an alignment occurring randomly within 1 arcsecond of Cyg OB2 \#5
is very low. Extragalactic source counts at 1.4~GHz indicate
$\sim70$~sources~deg$^{-2}$ of around 1~mJy
\citep[e.g.][]{Jackson:2005}, implying $\sim10^{-6}$ of these sources
in 1~arcsecond$^2$. If the source is at a much greater distance than
\object{Cyg OB2 \#5} it will have a different proper motion to the
binary. A distant galaxy will remain fixed relative to the reference
frame of background quasars while the binary and Star D will exhibit
proper motion relative to the frame. At radio wavelengths, it may be
possible to measure this proper motion through astrometry. Certainly,
the proper motion determined from the VLA observations discussed here
is less than 100 mas, and to improve this precision would require VLBI
observations. The low radio brightness of NE presents a challenge for
higher precision VLBI astrometry.  Alternatively, IR/optical imaging
could reveal if NE is associated directly with an object. \object{Cyg
OB2 \#5} is bright in the optical (V=9.21) and attempts at imaging the
region between the binary system and Star D are thwarted by the high
contrast of the binary and the large PSF of the imaging
telescopes. The IR image (Sec.~\ref{sec:irobs}) was searched for
evidence of a background source at the location of NE, but with no
success.  A smaller PSF could be attained by interferometry, but the
contrast could only be defeated through either adaptive nulling or
coronographic imaging. This observing challenge remains to be
attempted.

\section{Summary}
This paper re-examines over 50 VLA observations of the well-known
O-star binary system Cyg OB2\#5 in an attempt to locate and
characterize the well-known variable radio emission in the system.
The radio emission consists of a primary component that is associated
with the binary system and a component to the NE. Both components are
resolved in all 23 epochs of highest resolution VLA A-configuration
observations, which reveal the flux of NE is constant while the flux
of the primary varies. The constant flux from NE permits the flux of
the primary to be derived in all observations for the first time, most
especially those where the two components are not resolved
individually. A string-length analysis of the derived radio light
curves of the primary emission at both 4.8~GHz and 8.4~GHz gives a
period of $6.7\pm0.3$~years for the variations.

The primary emission changes character as it varies between its high
state of $\sim8$~mJy at 4.8 GHz and a spectral index of $0.24\pm0.01$
to a low emission state with a 4.8 GHz flux of $\sim$2 mJy and a
spectral index of $0.60\pm0.04$, consistent with thermal emission from
a stellar wind. Observations at 350 GHz obtained at an epoch not
during a low emission state are also consistent with the thermal
emission level observed during the low state, and hence lend support
to the argument that the thermal emission observed during the low
state remains relatively constant through the orbit.

The mass-loss rate of the binary is deduced to be
$3.4\times10^{-5}$~M$_\odot$~yr$^{-1}$ from the flux during the low
emission state, which is unusually high for an Of supergiant, and
comparable to the rates determined for WR stars. Together with the
anomalous strength of the He\,{\sc i} 1.083-$\mu$m emission line, also
redolent of a WR star, this points to a fast, heavy wind and supports
the suggestion by \cite{Bohannan:1976} that \object{Cyg OB2 \#5} is an
immediate progenitor of a WR binary system in which the mass loss has
not yet revealed enough evolved core material to affect the observed
spectrum.

The flatter spectral index during the high state is attributed to the
addition of a non-thermal component to the thermal emission from the
binary stellar wind. A non-thermal source orbiting within the
stellar wind envelope of the binary system every 6.7 years can account
for the variations in radio flux through orbit modulation of the
free-free opacity along lines of sight to the non-thermal source.

Such a model requires the presence of a third star in association with
the binary system. The high luminosity of the binary components and
the emission from the stellar wind make the detection of a third star
challenging. An analysis of radial velocity data from the literature
provides supporting evidence of reflex motion in the binary as a
result of a third star, labelled Star C, with a mass of
$23^{+22}_{-14}$~M$_\odot$. Until further observations can be
obtained, especially at orbital phases where the radial velocities
change most dramatically, this provides a tantalizing piece of
evidence that \object{Cyg OB2 \#5} is tertiary system rather than a
binary. Additional support for a third star comes from a hard X-ray
component, that \cite{Linder:2009} suggest arises in a WCR
between the binary wind and that of a third massive star.

This study also re-examines the NE source and its previous
identification with a WCR between the winds of the O-star
binary and that of a B0 star (Star D) $0.9\arcsec$ to the NE. Using
MERLIN observations at 1.4 GHz with the VLA observations confirms the
non-thermal nature of NE with a spectral index of $-0.50\pm0.12$, and
gives the relative separation of the binary and NE to be
$0.77\pm0.08\arcsec$. Higher frequency radio observations would
be useful to reveal the properties of the underlying relativistic
electron population. Through wind-momentum balance, the mass-loss
rate of Star D is estimated to be between
$0.5-6.1\times10^{-6}$~M$_\odot$~yr$^{-1}$, consistent with a
late-O/early B supergiant at the lower extremum of this mass-loss rate
range, and an order of magnitude too high for a lower luminosity
star. This raises the possibility that NE is not a WCR, though the
WR+B binary WR\,147 provides an example of an early B-type dwarf
providing a sufficiently dense wind to give a readily observed WCR
with the dense wind of a WN8 companion. Analysis of IR observations of
Star D to the NE reveal an apparent discrepancy between visual and
visual/IR-based reddening estimates, but provide no compelling reason
to rule out an association between \object{Cyg OB2 \#5} and Star D,
and hence the possibility that \object{Cyg OB2 \#5} is a quadruple
system. An estimate of the non-thermal luminosity of NE is also
consistent with a WCR. To test the alternative possibility that NE is
an unassociated background source requires either high precision
proper motion observations through VLBI, or deep optical IR
imaging. Both of these possibilities require very challenging
observations, that remain to be attempted.

\acknowledgments The authors would like to thank Julian Pittard,
Gregor Rauw, Mark Runacres, and Sven Van Loo for many useful
discussions related to this work.  This paper made extensive use of
data from the National Radio Astronomy Observatory Very Large Array,
New Mexico, USA, and the MERLIN array in England, UK. Thanks to Meri
Stanley and the analysts team at NRAO and to Anita Richards at MERLIN
for their help with the archive data. The National Radio Astronomy
Observatory is a facility of the National Science Foundation operated
under cooperative agreement by Associated Universities, Inc. MERLIN is
a National Facility operated by the University of Manchester at  
Jodrell Bank Observatory on behalf of the Science and Technology
Facilities Council (STFC) of the United Kingdom. The United Kingdom
Infrared Telescope is operated by the Joint Astronomy Centre for the
STFC.The James Clerk Maxwell Telescope is operated by The Joint
Astronomy Centre on behalf of the STFC, the Netherlands Organisation
for Scientific Research, and the National Research Council of Canada.
The Carlos S\'{a}nchez Telescope (TCS) of the Observatorio del Teide
(Tenerife) is operated by the Instituto Astrof\'{\i}sica de Canarias.
Facilities: \facility{VLA}, \facility{MERLIN}, \facility{JCMT(Scuba)},
\facility{UKIRT(IRCAM3), \facility{TCS}}                              


\end{document}